# Scaling up Action Through Collective Engagement with Environmental Data

Aksel Biørn-Hansen[1]

[1]*KTH Royal Institute of Technology, Stockholm, Sweden*

**Abstract**

Sustainability has over the past two decades emerged as a key concern in human-computer interaction, with a much critiqued focus on quantification and eco-feedback. This approach fits within a modernist framing of sustainability, treating the environment (and our impact on it) as an externality, reducing it to a set of simple metrics. While data about the climate impact of our actions provide an important indication of harm, such data is fragmented and incomplete, capturing only a partial picture of a very wicked and entangled problem. My doctoral research departs from this notion of "information will solve the problem" and through design-oriented explorations of environmental data such as $CO_2$ emissions from academic flying, I investigate alternative ways to engage people with environmental data in order to unsettle relations to the climate impact of our actions and foster care. So far, I have studied this through design-oriented case studies of data in action, with a specific focus on interventions aimed at engaging people in social contexts with the carbon emissions of everyday practices.

**Keywords**
Sustainability, Carbon Emissions, Environmental Data, Tangible Computing

## 1. Introduction

I am a third-year doctoral student in Human-Computer Interaction with a specialisation in sustainability at KTH Royal Institute of Technology, Sweden. The doctoral program is five years long, with a mix of research, coursework and teaching, all at once. In other words, I have two years left to figure out what I am doing. My supervisors are Daniel Pargman (main) and Elina Eriksson.

I have a strong interest in sustainability. Sustainability can mean different things to different people - for me, it encompasses the very urgent need for transformative changes to an inherently violent system built on capitalism, causing environmental destruction, extreme inequalities and injustices to both humans and more-than-humans alike. This necessitates not only a radical shift in e.g. how society operates and use resources, but also, and maybe more importantly, in the underlying values and relations on which the systems rests on. With this in mind, you could say that I embrace a radical view of sustainability, rejecting a position where technology would be the primary solution.

Despite the wealth of knowledge we have about the human impact on the planet, far too little is being done to change course and make the necessary, system-wide changes needed to

---

move toward a low-carbon future. A key obstacle is the challenge of translating knowledge into action [1]. The complexity, temporality and situated nature of the climate crisis makes it hard to comprehend and act upon. This is further clouded by the abstract nature of the large amounts of environmental data we collect and measure in order to make sense of cause and effect. While data about the climate impact of our actions provide an important indication of harm, such data is fragmented and incomplete, capturing only a partial picture of a very wicked and entangled problem. Data is furthermore abstract and intangible, which makes it hard to engage with as it lacks material presence, accumulating without being meaningfully used [2]. Environmental data is particularly difficult given the invisibility of e.g. carbon emissions, which for the most part are far removed from people's everyday lives.

Building on research that challenges the focus on reductive design interventions within sustainable human-computer interaction [e.g. 3, 4], my doctoral research aims to dig deeper into the challenge of supporting action towards sustainability at scale through design-oriented explorations of environmental data such as $CO_2$ emissions. I do so through investigating alternative ways to engage people with environmental data as a means to facilitate new conversations, meaning-making and action toward decreased $CO_2$ emissions and increased sustainability. So far, I have studied this through two case studies of data in action, with a specific focus on interventions aimed at engaging people in social contexts with the carbon emissions of everyday practices. In what follows, I will briefly describing in more detail my research approach and the results so far, before ending the paper by discussing possible shortcomings and next steps for my dissertation.

## 2. Research approach

The main objective of my doctoral research is to contribute with knowledge and practical tools to support a transition toward low-carbon futures. As described in the introduction I have worked towards this goal through design-oriented explorations of environmental data and alternative ways to engage people with environmental data as a means to facilitate new conversations, meaning-making and action toward decreased $CO_2$ emissions and increased sustainability. In this work, I apply a design-oriented research approach [5], meaning that I use design methods and the design of artefacts to create new knowledge. The key rational behind this choice is that through design, I can iteratively create and critique potential responses to a problem, and through doing so continuously reframe the problem at hand in order to arrive upon "a concrete problem framing and articulation of the preferred state, and a series of artifacts–models, prototypes, products and documentation of the design process" [6, p. 497]. Additionally, I have in my research so far made use of interviews, workshops and observations as methods to study how people interact and use technology. However, given the interdisciplinary nature of my research, I also dip into other methods and epistemologies beyond qualitative research, and will continue to do so on my journey towards a more integrative and undisciplinary [7] research approach.

Through two case studies I hope to contribute with insights about how to design for sustainability [8] and to the general sustainability discourse, offering directions for how to approach the goal of emission reductions from a more holistic framing and relation to sustainability.

## 3. Results so far

### 3.1. Engaging people with environmental data

The first case study explores ways to engage people with environmental data. As part of the research project «Decreased $CO_2$ emissions in flight-intensive organisation: from data to practice» (FLIGHT), I have together with colleagues designed and studied different kinds of data representations and ways to engage employees at our own university with data on the climate impact of business air travel. The project has access to a large data set containing data on who flies where and when at KTH between 2017-2021. Through unpacking this data set and exploring different ways to represent and invite people to make sense of it as part of workshops, the aim has been to support and facilitate conversations on how shared practices need to change in order to reduce emissions and reach highly ambitious climate targets at KTH.

This work began in 2020 with a first study of the available data set and different interactive visualizations created in collaboration with students at KTH. The visualizations were aimed at three different levels in the organisation: bottom-up, middle and top-down visualizations. Findings from this study are published in a journal paper [9] and highlights the middle level, the level in-between the individual and the larger organisation, i.e. departments, as an opportune place to intervene when the purpose is to trigger and mediate conversations around emission reductions and work practices using data. Moreover, this study discusses several complications concerning data and ethics, such as the structural changes needed to enable proper follow-up of the $CO_2$ emissions from academic flying as well as the thorny issue of where the border goes between shaming individuals versus constructively supporting action.

As an extension of the above, I have investigated more in-depth the use of data physicalisation and the materialization of $CO_2$ emissions as a means to engage people with environmental data. Adopting the sentiment proposed by Pierce and Paulos [10] who suggests that we should shift from "shouting at people about energy to inviting them to be more in touch with energy" [10, p. 121], the aim with this specific investigation has been to support more embodied sense-making processes in a participatory and non-prescriptive manner through interactions with the materiality of $CO_2$ emissions from flying. For these purposes, my colleagues and I have designed a low-tech data physicalisation involving poker chips and post-its spread out on a grid of squares, drawing on literature around materialisation and physicalisation [11, 12, 13, 14, 15, 16, 17].

The complete board represents one department and each square represents one anonymous employee at the department on which stacks of chips are placed, representing the number of flights and corresponding $CO_2$ emissions (see Figure 1). Over a period of three years between 2019 and 2020, we iterated on this design and also used it as part of workshops with different departments at KTH where we presented employees with their own flying data. This whole process provided many insights into the design of data physicalisation aimed at supporting a collective meaning-making process amongst different groups of people. It also showed how this method of unpacking environmental data could scale or move across different levels of the organisation, acting as a gateway to broader discussions about reducing emissions from business air travel in the organisation amongst different actors. In a forthcoming paper submitted to DIS2023 [1] we describe these insights in more detail and raise a set of methodological considerations for data physicalisation aimed at supporting new conversations, meaning-making and actions toward decreased $CO_2$ emissions and increased sustainability.

---

[1]https://dis.acm.org/2023/

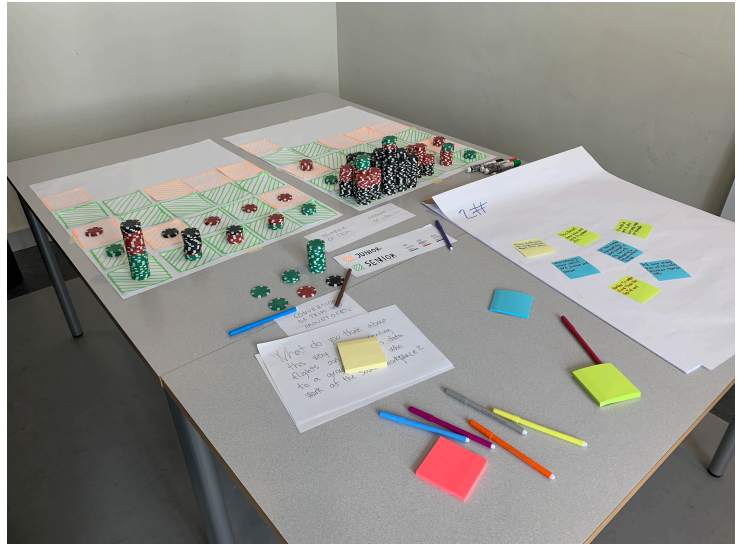

**Figure 1:** The data physicalisation in its final iteration, consisting of two boards with coloured squares representing anonymous employees at a division with stacks of poker chips representing number of flights (left) and amount of $CO_2$ emissions (right). Green poker chips represent short-haul flights, red represents medium-haul flights and black represent long-haul flights.

### 3.2. The use of a carbon calculator in social settings

In parallel with the work in the FLIGHT project, I have also taken part in the now completed research project Habitwise. The overall aim of the project has been to investigate how organisations can support their employees to adopt more sustainable everyday habits. This has been investigated mainly through the introduction of longitudinal workplace campaigns at two companies in the form of what you might call step competitions, but with a focus on saving $CO_2$ emissions instead. As part of these campaigns, a novel carbon calculator was used to track individual carbon footprints and performance amongst participants. Historically, carbon calculators have been focused on "only" communicating the carbon footprint of individuals, leaving users to figure out how to make choices supporting a decreased climate impact. The carbon calculator used in this project, called Habits, contains additional gamified features such as concrete challenges for reducing the carbon footprint as well as social features meant to give support beyond pure information provision. Roughly explained, the campaigns were "conducted" during 2021 and lasted for six to eight weeks in which teams of employees at these two companies competed on who could save the most $CO_2$ emissions possible using the calculator and the challenges it provided as shared interface.

In this project I have together with colleagues conducted two studies, one on the use of this carbon calculator "in the wild" and a large study of these campaigns explained above. The first study was an interview study conducted in 2020 with a group of people representing users in search of a climate calculator, i.e. who discovered and used Habits "in the wild". The aim was to understand their experience of Habits: Their discovery of it on the internet, motivations for using it, their interaction with it, and their assessment of how it met their expectations. The results from this study were presented in a paper at NordiCHI2022 [18] and highlight a set of challenges as well as opportunities for design, such as how the lack of social context and adaptive design put barriers to engagement beyond first time use. The results also point toward a general ambivalence or limitation with this kind of tool in supporting a change in behaviour, and point out an opportunity to support action beyond awareness.

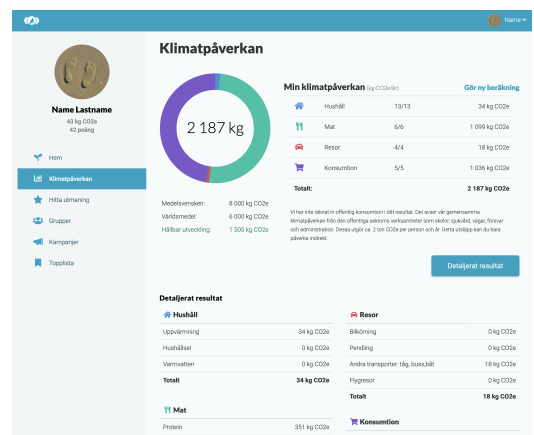

**Figure 2:** Screenshot of the result from the calculation of the carbon footprint in *Habits*. On the left is the sidebar menu, and in right centre is a overview and breakdown of the users carbon footprint across four areas: housing, food, travel and consumption.

The second study is the study of the two campaigns described above. For this we gathered large amounts of data using a mixed methods approach involving multiple surveys, usage data as well as follow-up interviews with campaign participants after the campaigns were finished. Currently we are in the process of analysing and writing up the results from this process. Preliminary results show that those with a high performance also had a strong sense of being part of a team, which the interviews corroborate as well, highlighting the importance of a social context or space to ground interactions with this kind of carbon data. In some sense we can say that the carbon calculator in this context acted as an interface in-between individuals and teams, between employees and the company.

## 4. Discussion

In my research so far, I have been able to synthesise many different kinds of data streams. Findings from this work highlight insights related to data, design and scale. However, the focus on how to make data actionable and the various studies done so far does not enable me to say anything meaningful about what actually makes data actionable. I have so far not studied the effects of exposure to the different data representations and what happened after interacting with the data in the different projects. I see this a critical limitation if the framing of my doctoral research is supposed to focus on action. A consequence of this could be to reframe my work to focus on meaning-making, sense-making or collaborative engagement with environmental data in some shape or form, ditching the action all together. Another possibility could be to redefine actionable as something that empower people to form new perspectives and relations to the impact of their action, e.g. flying, and what kind of experiences and thoughts that results in close proximity to the studied data or data artefact. Another important limitation in that I have not properly defined or investigated what environmental data really is. To address this, it would be necessary to more closely conceptualise environmental data, its characteristics, its sources, different actors involved in production and use, power dynamics etcetera. Furthermore, environmental data such as data on the climate impact of academic flying, reveals the symptoms of a harmful, violent system of pollution. Through understanding what environmental data is, it could help me come closer to the systems that produces this data and also allow me to scale the focus to organisational or infrastructural change. For now, I will leave it as a possible future work, but it is definitely an important pieces of my work that I need to think more about.

## 4.1. Sustainability

Up until very recently, my research has focused a lot on amounts of $CO_2$ emissions. This is in line with modern conceptualisations of sustainability as an reductive concept focused on "the race to the bottom" and purity activism [19]. It is important to reduce emissions and transition toward a low-carbon society. However, this is a very limited view of sustainability and focuses on the symptoms rather than the causes of the emissions in the first place. This sort of scalar missmatch [19] conflates relationships that operate at different scales, asking questions of "how much" instead of "how" and "why" questions that can be argued to be inherently more important. Considering this perspective and past critiques of SHCI [e.g. 3], a major challenge for me going forward is to try to bridge this question of scale and integrate a more complex and nuanced perspective of sustainability into my work. How I will do this is still a bit unclear to me, but a first step will be to harness the data I have gathered in the FLIGHT project so far and unpack the complexity of academic flying in relation to data. This will not require more data collection, but rather a synthesis of existing data as well as analysis of it in relation to this perspectives of scale in sustainability.

## 4.2. Participation in the ICT4S doctoral symposium 2023

Through participating in this doctoral symposium, I hope to get some critical feedback on this work, especially on the overarching framing of it, in order to develop it further. I still have time to develop my research in new directions. I can also confess that I currently have this sort of "halfway-PhD-depression", being a bit tired of my own work, and can do with a bit of inspiration. Taking part in the research of other doctoral students and give some feedback in return will be helpful.

# References

[1] A. Kollmuss, J. Agyeman, Mind the gap: why do people act environmentally and what are the barriers to pro-environmental behavior?, Environmental education research 8 (2002) 239–260.

[2] W. T. Odom, A. J. Sellen, R. Banks, D. S. Kirk, T. Regan, M. Selby, J. L. Forlizzi, J. Zimmerman, Designing for slowness, anticipation and re-visitation: A long term field study of the photobox, in: Proceedings of the SIGCHI Conference on Human Factors in Computing Systems, CHI '14, Association for Computing Machinery, New York, NY, USA, 2014, p. 1961–1970. URL: https://doi.org/10.1145/2556288.2557178. doi:10.1145/2556288.2557178.

[3] H. Brynjarsdóttir, M. Håkansson, J. Pierce, E. P. Baumer, C. DiSalvo, P. Sengers, Sustainably unpersuaded: How persuasion narrows our vision of sustainability, in: Conference on Human Factors in Computing Systems - Proceedings, ACM Press, New York, New York, USA, 2012, pp. 947–956.

[4] P. Dourish, HCI and environmental sustainability: The politics of design and the design of politics, in: DIS 2010 - Proceedings of the 8th ACM Conference on Designing Interactive Systems, ACM Press, New York, New York, USA, 2010, pp. 1–10. URL: http://portal.acm.org/citation.cfm?doid=1858171.1858173. doi:10.1145/1858171.1858173.

[5] D. Fallman, Design-Oriented Human-Computer Interaction, in: Proceedings of the SIGCHI Conference on Human Factors in Computing Systems, CHI '03, Association for Computing Machinery, New York, NY, USA, 2003, pp. 225–232. URL: https://doi.org/10.1145/642611.642652. doi:10.1145/642611.642652, event-place: Ft. Lauderdale, Florida, USA.

[6] J. Zimmerman, J. Forlizzi, S. Evenson, Research through design as a method for interaction design research in HCI, in: Proceedings of the SIGCHI Conference on Human Factors in Computing Systems - CHI '07, ACM Press, San Jose, California, USA, 2007, pp. 493–502. URL: http://dl.acm.org/citation.cfm?doid=1240624.1240704. doi:10.1145/1240624.1240704.

[7] L. J. Haider, J. Hentati-Sundberg, M. Giusti, J. Goodness, M. Hamann, V. A. Masterson, M. Meacham, A. Merrie, D. Ospina, C. Schill, et al., The undisciplinary journey: early-career perspectives in sustainability science, Sustainability science 13 (2018) 191–204.

[8] J. C. Mankoff, E. Blevis, A. Borning, B. Friedman, S. R. Fussell, J. Hasbrouck, A. Woodruff, P. Sengers, Environmental Sustainability and Interaction, in: CHI '07 Extended Abstracts on Human Factors in Computing Systems, ACM, New York, NY, USA, 2007, pp. 2121–2124. URL: http://www.pigeonblog.mapyourcity.net/. doi:10.1145/1240866.

[9] A. Biørn-Hansen, D. Pargman, E. Eriksson, M. Romero, J. Laaksolahti, M. Robért, Exploring the problem space of co2 emission reductions from academic flying, Sustainability 13 (2021) 12206.

[10] J. Pierce, E. Paulos, Materializing Energy, in: Proceedings of the 8th ACM Conference on Designing Interactive Systems - DIS '10, ACM Press, New York, New York, USA, 2010, pp. 113–122. doi:10.1145/1858171.

[11] Y. Jansen, P. Dragicevic, P. Isenberg, J. Alexander, A. Karnik, J. Kildal, S. Subramanian, K. Hornbæk, Opportunities and challenges for data physicalization, in: Conference on Human Factors in Computing Systems - Proceedings, Association for Computing Machinery, Seoul, Korea, 2015, pp. 3227–3236. URL: http://dx.doi.org/10.1145/2702123.

2702180. doi:10.1145/2702123.2702180.

[12] S. Huron, S. Carpendale, A. Thudt, A. Tang, M. Mauerer, Constructive visualization, in: Proceedings of the 2014 Conference on Designing Interactive Systems, DIS '14, Association for Computing Machinery, New York, NY, USA, 2014, p. 433–442. URL: https://doi.org/10.1145/2598510.2598566. doi:10.1145/2598510.2598566.

[13] R. Soden, N. Kauffman, Infrastructuring the imaginary how sea-level rise comes to matter in the San Francisco Bay area, in: Conference on Human Factors in Computing Systems - Proceedings, ACM, Glasgow, Scotland, UK, 2019, pp. 1–11.

[14] R. Soden, P. Hamel, D. Lallemant, J. Pierce, The disaster and climate change artathon: Staging art/science collaborations in crisis informatics, in: DIS 2020 - Proceedings of the 2020 ACM Designing Interactive Systems Conference, ACM, Eindhoven, Netherlands, 2020, pp. 1273–1286.

[15] B. Stegers, K. Sauvé, S. Houben, Ecorbis: A Data Sculpture of Environmental Behavior in the Home Context, in: Designing Interactive Systems Conference, ACM, Virtual Event, Australia, 2022, pp. 1669–1683. URL: https://doi.org/10.1145/3532106.3533508. doi:10.1145/3532106.

[16] T. De Greve, S. Malliet, N. Hendriks, B. Zaman, The Air Quality Lens: Ambiguity as Opportunity to Reactivate Environmental Data, in: Designing Interactive Systems Conference, ACM, Virtual Event, Australia, 2022, pp. 335–348. URL: https://dl.acm.org/doi/10.1145/3532106.3533530. doi:10.1145/3532106.3533530.

[17] D. Holstius, J. Kembel, A. Hurst, P.-H. Wan, J. Forlizzi, Infotropism: living and robotic plants as interactive displays, in: Proceedings of the 5th conference on Designing interactive systems, ACM, Cambridge, Massachusetts, USA, 2004, pp. 215–221.

[18] A. Biørn-Hansen, C. Katzeff, E. Eriksson, Exploring the use of a carbon footprint calculator challenging everyday habits, in: Nordic Human-Computer Interaction Conference, NordiCHI '22, Association for Computing Machinery, New York, NY, USA, 2022. URL: https://doi.org/10.1145/3546155.3546668. doi:10.1145/3546155.3546668.

[19] M. Liboiron, Pollution is colonialism, Duke University Press, Durham, 2021.